\documentstyle[twoside,fleqn,espcrc2,epsf]{article}

\def\inseps#1#2{\def\epsfsize##1##2{#2##1} \centerline{\epsfbox{#1}}}

\newcommand{\AmS}{{\protect\the\textfont2
  A\kern-.1667em\lower.5ex\hbox{M}\kern-.125emS}}

\title{Critical behaviour of the three-dimensional gonihedric 
Ising Model}

\author{E.N.M. Cirillo$^{\rm a}$, G. Gonnella
         \address{Dipartimento di Fisica
         dell'Universit\`a degli Studi di Bari and
	 Istituto Nazionale di Fisica Nucleare, Sezione di Bari
         via G. Amendola 173, I-70126 Bari, Italy}
         ,
         A. Pelizzola
         \address{Dipartimento di Fisica del Politecnico di Torino and
	 Istituto Nazionale di Fisica per la Materia,
         c. Duca degli Abruzzi 24, 10129 Torino, Italy}}

\begin{document}

\begin{abstract}
We use the cluster variation method (CVM) and Monte Carlo
simulations to investigate the phase structure 
of the 3d gonihedric Ising actions defined by Savvidy and Wegner.
This model corresponds to the usual three-dimensional cubic
Ising model with  nearest,  next to the nearest, and plaquette interactions
in the region with degenerate lamellar ground states.
The picture of the phase diagram given by the CVM is
in good agreement
with the results of Monte Carlo simulations, and it is shown that the
gonihedric model is in the same universality class of the ordinary
three-dimensional Ising model.
\end{abstract}

\maketitle

The 3d Ising model with interactions extended to next-to-the-nearest
neighbouring (NNN) sites and 4-spin plaquette interaction
has a very rich phase diagram relevant for the description 
of physical systems of interacting interfaces \cite{1,2}.
In this paper we discuss the critical behaviour 
of this model in a particular region of parameters first
explicitly considered in \cite{3} in relation with 
string theory.

Ising variables can describe surface configurations in a
simple way to explain: for any spin configuration
there is on the dual lattice a set of closed interfaces
separating the domains of spins with different sign.
If $\beta_A, \beta_C, \beta_I $ respectively are  the energy
cost for an unit area, two connected plaquettes forming a right
angle, and four plaquettes intersecting at a common link, the corresponding
Ising model is given by 
\begin{equation}
\begin{array}{lcl}
{\bf -}H &=& J_1 \sum_{\langle ij\rangle }\sigma_{i} \sigma_{j} +
J_2 \sum_{\langle \langle i,j\rangle \rangle }\sigma_{i} \sigma_{j}\\ 
&+& J_4 \sum_{[i,j,k,l]}\sigma_{i} \sigma_{j}\sigma_{k} \sigma_{l}.
\end{array}
\label{e1}
\end{equation}
with $J_1=\beta_C +(\beta_A + \beta_I)/2, J_2 = -(2 \beta_C + \beta_I)/8,
J_4 = (2 \beta_C - \beta_I)/8$ \cite{4}. 
In eq.\ (\ref{e1}) the three sums  respectively are on the nearest 
neighbours, the NNN spins and the plaquettes of the cubic lattice.
The gonihedric model is the model  (\ref{e1})
with $\beta_A=0$ $(J_2/J_1=-1/4)$ and $\beta_C=1$. The choice $\beta_A=0$
means that the amount of surfaces in the system is not controlled by any 
external parameter, while $\beta_C=1$ gives a preference to flat surfaces.
The parametrization 
$J_1=2 k, J_2 = -k/2, J_4 =(1-k)/2$ is often used; the parameter $k=\beta_I/4 + 1/2$
can interpolate between completely non-interacting ($k=1/2$)
and self-avoiding surfaces ($k=\infty$).

The ground state degeneracy is the characteristic of the gonihedric
Ising model:
all the possible sequences of ``+'' and ``-'' spin planes (for $J_4/J_1 > -1/4$)
have the same minimum energy. The situation changes for $\beta_A \neq 0$:
a positive $\beta_A$ selects the ferromagnetic ground state (no interface 
allowed) while  at negative $\beta_A$ the ground state is a sequence
of planes with spins of alternate sign (maximum amount of flat interfaces).

The finite temperature behaviour of the model can be easily studied by
means of the mean field approximation, which predicts that the ground
state symmetry is preserved at finite temperature (that is, the first
order line between lamellar and ferromagnetic phases in Fig. 1 would be
strictly vertical, independent of temperature) and, at least for $k >
3/4$, the ferromagnetic and lamellar phases are separated from the
disordered one by two second order lines (both turning first order for
$k < 3/4$) which meet in a bicritical point at $J_2/J_1 = -1/4$.

Recently, several studies based on high
temperature expansions \cite{5} and Monte Carlo simulations \cite{6}
addressed the issue of the
universality class of the gonihedric model (the existence of a phase 
with long-range order has been analytically proved in \cite{7}
for the case $J_4=0$), formulating the conjecture that the model might
be in the Onsager (i.e. two-dimensional Ising) universality class.
Further, longer simulations \cite{8}, however, found estimates for
critical exponents which were seriously incompatible with both the
conjecture and the previous numerical estimates. 

We have studied the gonihedric model at finite temperature
by means of the cluster variation method (CVM) \cite{9,10}, which is a powerful 
generalization of the mean field approximation \cite{11}. 
Our results show that the ground-state degeneracy
along the line $J_2/J_1=-1/4$ is broken at finite temperatures and the 
ferromagnetic phase is stable along this line.
In Fig. 1 we show the phase
diagram in the space $(1/(\beta J_1), J_2/J_1 (J_4=0)$ 
($\beta=(k_B T)^{-1}$) as calculated in the cube approximation of the CVM.
We see that
the ferromagnetic-lamellar transition line bends at finite temperatures
towards values of $J_2/J_1$ lower than -1/4;
this bending has been confirmed by low-temperature 
expansion results \cite{9,12} and is very important since, with the
reasonable assumption that 
the ferromagnetic-paramagnetic line, which is of second order,
does not exhibit non-universal behaviour, we see that the gonihedric
model must be in the same universality class of the ordinary
three-dimensional Ising model.
 
The lamellar to disordered transition line turns out to be first order
for $J_4 = 0$ 
(this being indipendently confirmed by the numerical simulations in
\cite{13}), and hence the correct phase diagram exhibit a critical end point
structure, instead of the bicritical one predicted by the mean field
approximation. Nevertheless, CVM results for the critical amplitude
of the Ising transition close to $J_2/J_1=-1/4$,
analyzed with Pad\`e approximants \cite{14} show the existence of strong
cross-over effects \cite{10}, probably due to the vicinity of the
lamellar spinodal line, which should be responsible  
for the contrasting results
for the critical exponents found in the simulations
\cite{6,8}.  Other  previous simulations \cite{13}
were also not able to give  definitive estimates of critical exponents.

At $J_2/J_1=-1/4$
the critical value of $1/(\beta J_1)$ is $1.171$.
It can  also be shown by CVM calculations that for other values
of $J_4$ the ferromagnetic-paramagnetic transition becomes 
of first order close to the lamellar transition with a tricritical
point at $J_2 < 0$. In the $k$-parametrization 
the tricritical point is observed
for $k < 0.86$.

\begin{figure}[htb]
\vskip -25mm
\inseps{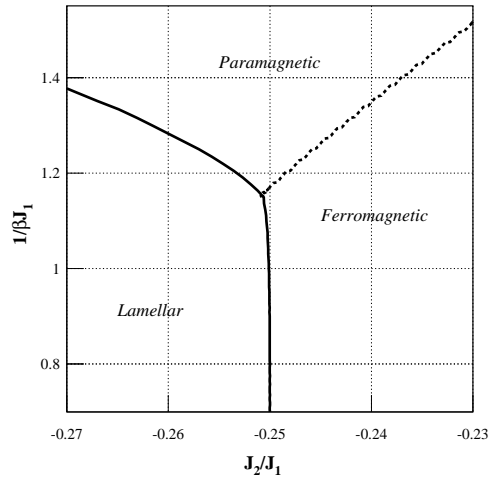}{0.45}
\vskip -30mm
\caption{The phase diagram of model (\ref{e1}) at $J_4=0$. Solid and
dashed lines represent first and second order transitions respectively.}
\end{figure}

 
In order to understand the source of the discrepancies between the
various simulations, we tried also to perform a standard Metropolis
Monte Carlo on this model.
In Fig. 2 we give our results for the behaviour of the specific
heat. Our data have been obtained
by averaging over $2000$ decorrelated measures at each value
of the lattice size $L$ and of the inverse temperature $\beta$. 
Our decorrelation times were $1000$ full updates of the lattice
in the case of the largest value of $L$. These times are substantially
longer than those reported in \cite{6,8}, but we found them necessary
(at least with $L = 18$) in
order to get stable averages. This suggests that the simulations
performed in \cite{8} with much larger lattices are definitely too
short.

We fitted our data with the function
$C(\beta)=C_{\rm max}/[b(\beta-\beta_c)^2+1]$ and  
the values of $C_{\rm max}$ and $\beta_c$ obtained for
each lattice size are listed in Table 1.
A fit of the position of the peak with the function $\beta_c=\beta_{\rm crit}
+aL^{-\frac{1}{\nu}}$ gives the
estimate for the
critical inverse temperature $\beta_{\rm crit}=0.4370\pm 0.0001$
corresponding to $1/(\beta J_1)=1.144$ to be compared
with the CVM value $1.171$ and with $1.136$   
of  the previous simulations \cite{6}.
Our estimate for the exponent
$1/\nu$ is
$1/\nu=1.483\pm 0.001$. Since the Ising value is $1/\nu=1.594\pm 0.004$ 
\cite{15} we have to go to much larger lattices in order to get reliable
estimates. 

\begin{table}[hbt]
\begin{tabular}{ccc}
\hline\hline
$L$    &    $C_{\rm max}$    &   $\beta_c$   \\
\hline\hline
$8$      &   $2.8041\pm 0.0237$  &  $0.41182\pm 0.00021$ \\ 
$10$     &   $4.3707\pm 0.0706$  &  $0.41872\pm 0.00018$ \\
$12$     &   $6.1137\pm 0.0334$  &  $0.42300\pm 0.00010$ \\
$16$     &   $9.4068\pm 0.0586$  &  $0.42794\pm 0.00004$ \\
$18$     &  $10.6710\pm 0.0883$  &  $0.42937\pm 0.00002$ \\
\hline\hline
\end{tabular}
\vskip 0.5 cm
\caption{Valus of $C_{\rm max}$ and $\beta_c$ for different choices
of the lattice size $L$ obtained by fitting data in Fig. 2.}
\label{tab}
\end{table}

\begin{figure}[htb]
\vskip -35mm
\inseps{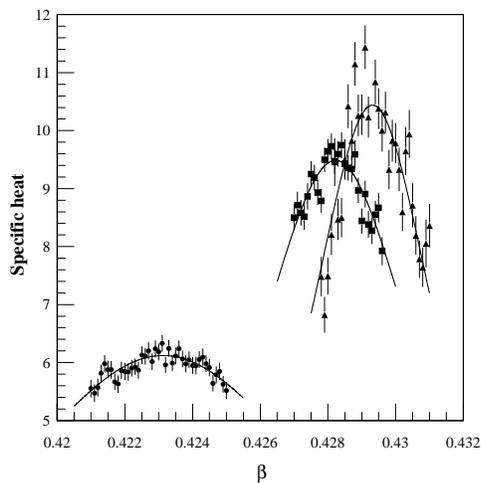}{0.45}
\vskip -30mm
\caption{Monte Carlo measurements of specific heat in the gonihedric
model with $\kappa=1$. Circles, squares and
triangles refer respectively to $L=12,16,18$. Solid lines represent the best
fit of the peaks performed with the function given in the main text.}
\end{figure}



In summary, we have shown, studying the phase diagram of the model in an
enlarged parameter space by means of the cluster variation method,
that the three dimensional gonihedric model is in the same
universality class of the ordinary three dimensional Ising model.
Furthermore, the model seems to exhibit very strong crossover effects
which pose serious problems to numerical simulations.

We are indebted to Marcia Barbosa, Amos Maritan, Michele Caselle,
Dino Cosmai and Paolo Cea for helpful discussions about
this work.


\end{document}